# Application of the Ranking Relative Principal Component Attributes Network Model (REL-PCANet) for the Inclusive Development Index Estimation

**Anwar Irmatov**
Lomonosov Moscow State University,
Russian Federation
irmatov@intsys.msu.ru

**Elnura Irmatova**
Zhejiang University,
China
Irmatova.17@intl.zju.edu.cn

## Abstract

In 2018, at the World Economic Forum in Davos it was presented a new countries' economic performance metric named the Inclusive Development Index (IDI) composed of 12 indicators. The new metric implies that countries might need to realize structural reforms for improving both economic expansion and social inclusion performance. That is why, it is vital for the IDI calculation method to have strong statistical and mathematical basis, so that results are accurate and transparent for public purposes. In the current work, we propose a novel approach for the IDI estimation - the Ranking Relative Principal Component Attributes Network Model (REL-PCANet). The model is based on RELARM and RankNet principles and combines elements of PCA, techniques applied in image recognition and learning to rank mechanisms. Also, we define a new approach for estimation of target probabilities matrix $T_{Rnet}$ to reflect dynamic changes in countries' inclusive development. Empirical study proved that REL-PCANet ensures reliable and robust scores and rankings, thus is recommended for practical implementation.

# CONTENTS

# I. Introduction

The economic growth model that worked in recent decades now changing to become socially inclusive instead of basing only on supply, export and private capital investment. Globalization and fast technological change led to strengthening inequality and almost unchanging median income.

In current times GDP growth is still the key country performance indicator although the world starts to understand that the more attention should be payed to socioeconomic progress in economic policy.

In 2018, at the World Economic Forum in Davos it was presented a new performance metric named the Inclusive Development Index within Shaping the Future of Economic Progress framework. WEF report showed that GDP growth is not sufficient for reaching higher levels of living standards and what is more majority of citizens do not assess countries' economic efficiency by GDP but by group of indicators describing household's standard of living. Moreover, steady and comprehensive progress accompanied by the growth of incomes of the population and growth of its economic opportunities as well as the level of security and quality of life should be the priorities of the new economic policy and the main goal of economic development but not GDP growth. This created the prerequisites for the formation of a new assessment tool of economic development effectiveness – Inclusive Development Index (IDI) which is calculated for 107 countries and includes 3 blocks (12 indicators): Growth & development, Inclusion and Intergenerational equity & sustainability.

*Problem formulation*

The new economic metric implies for countries that they should implement structural reforms in order to realize changes for improving both economic expansion and social inclusion performance. Such kind of transformations might require enhancement of the overall country's ecosystem as well as restructuration of policies intended to improve living standards. That is why the calculation method of the Inclusive Development Index is very important and for its application to public purposes it should be accurate and transparent for all its possible users.

At the moment, IDI calculation method is based on linear min-max transformation which can lead to significant bias or smoothing of the final results. Also, this type of computation does not take into account whole interdependencies between countries which are crucial as economic development should be assessed not in isolated system but taking into whole countries' interactions. Here, we narrow this problem to finding an appropriate combination of rating and learning to rank models.

In current work we suggest a new model for IDI estimation named the Ranking Relative Principal Component Attributes Network Model (REL-PCANet) which employs principles of the rating model based on relative PCA attributes (RELARM) [1], Deep relative attributes [2] and RankNet [3].

Overall, REL-PCANet provides the following benefits for IDI estimation:

 1. Clarity of calculation methods and trustworthiness of the final results as REL-PCANet is constructed using statistical and economic modeling techniques with implementation of machine learning mechanisms. The aim of such combination is to achieve more reliable and precise ranking system.
 2. Ability to take into account comprehensive interdependencies between countries;
 3. Ability to build and train a model on a dataset with short time horizon.

Moreover, REL-PCANet reflects dynamic changes in countries' inclusive development due to special model of target probabilities matrix $T_{Rnet}$ proposed in this paper. Thus, proposed REL-PCANet reflects occurred economic and social changes as well as gives a possibility to make forecast for future index movements. Additionally, empirical study shows that REL-PCANet ensures robust results.

Next, we briefly describe existing rating and learning to rank models, then present a new approach for calculation of the Inclusive Development Index followed by an empirical example and draw the conclusions.

## II. Rating and learning to rank models

Rating models are essential instruments intended to help government, companies and households to simplify decision making process connected with choosing the right object based on large number of its features. Generally, rating models can be conditionally divided into 3 types: models using expert judgment, models using econometric tools, models using machine learning techniques.

The first type is not applicable to our problem because it contains large amount of expert component. Econometric method for model construction is widely used for credit rating assignment. The most commonly used form of regression in that field is a logistic regression (simple one, fuzzy [4] or ordinal [5]). However econometric model requires vast database for model training and adjustment, otherwise, there might occur substantial errors in final assessments that is why it is also not suitable for IDI.

Over the last decade machine-learning techniques such as neural networks and support vector machines were intensively developed. They have been widely used in image recognition systems as well as in theoretical rating modeling and learning to rank mechanisms. One can also find implementation of principal component analysis (PCA) in construction of rating models [6], [7], [8].

Learning to rank mechanisms are mainly applied in information retrieval, machine learning and natural language processing. Learning to rank can be described as implementation of machine learning methods for training model that solves ranking task. Lui in [9] divides learning to rank approaches in 3 groups: pointwise, pairwise and listwise. The point-wised method is the simplest one and practical implication showed that the other two perform better. The models using this method can be regression or ordinal regression based and classification based (ex. Pranking [10], MCRank[11]). The input to the training model for listwise method composes of a list of ranked objects so that the problem changes from ranking to optimization (ex. SoftRank[12], AdaRank[13], ListNet[14]) Finally, pairwise method uses pairs of objects where each of them has a specific label showing a relevance between them. The vivid examples of pairwise approach implementation are RankNet [3] and RankBoost [15]. Pairwise approach has similar ideas with the relative attributes principles that is why it is assumed in this paper to be compatible together.

## III. A new approach for IDI estimation - Ranking Relative Principal Component Attributes Network Model (REL-PCANet)

In current section we propose a new model for IDI computation named the Ranking Relative Principal Component Attributes Network Model. As it was mentioned in introduction, it is based on RELARM and RankNet principles and combines elements of PCA, techniques applied in image recognition and learning to rank mechanism which contains a neural network. Also, we propose a new technique for estimation of the network's target probabilities matrix which allows to reflect dynamic changes in countries' inclusive development. In this section we describe reasons for choosing RELARM and RankNet elements for constructing a model for IDI estimation, then provide brief description of their theoretical frameworks and finally present the new approach - REL-PCANet.

### *3.1 Implementation of RELARM and RankNet for REL-PCANet construction*

To begin with, the Relative Attributes Rating Model has the following distinctive features:

1. Application of specially defined relative PCA attributes, rating and ranking vectors and special ranking functions to rating/ranking purposes;
2. The use of minimum expert component which is limited to choice of initial model parameters;
3. Assessment of particular feature taking into account comprehensive objects' interdependencies. RELARM is based on the principle of "living organism" where each element change (even very small) causes certain reflection on the state of other analyzed system objects;
4. Simplicity of model training and calculation on small but relevant data array. RELARM can be trained on the 1-2 years data, so that it is becoming unnecessary to use large training samples.

In RELARM special role is given to the relative PCA attributes which provide the most comprehensive description of analyzed rating object characteristics. The concept of attributes is widely used in image recognition algorithms. It is most often presented in recognition using binary properties, which predicts a presence or an absence of a specific attribute (e.g. smiles on photos, determination of a landscape type etc.). However, the use of such algorithms has certain restrictions and often leads to ambiguous recognition or total disregard of a characteristic. Later in paper [16] it is proposed an application of relative attributes providing semantically more rich method for object description, which uses objects features comparison in relation to each other. The concept of relative attributes provides a relative strength of specified features presence of an object compared to other objects.

However, in the work [1] there was proposed a new way for relative attributes application combined with principal component analysis elements named Relative PCA Attributes. They are used in conjunction with the specially created rating vector along with ranking vector and function for obtaining relative PCA attribute ranking function values. Such a combination in the whole provides reliable and robust results and makes the RELARM indispensable for practical usage.

RELARM applies k-means clustering algorithm [17] for final rating assignment. Although for ranking purposes it might have been more suitable to calculate projections of relative attribute ranking function values to the rating vector and form a ranking. However, such an estimation might not reflect dynamic year to year changes or lead to performing undesirable outliers which is why in current work we enhance RELARM with adding a neural network mechanism for building rankings. We found that the deep relative attributes concept [2] works well with RELARM's underlying ideas, so that RankNet algorithm elements for ranking were assumed acceptable. RankNet was firstly presented in the article [3] and it was intended to use for information retrieval. Besides it found practical implications in various areas and especially in [2] where authors presented deep neural network with special ranking layer based on RankNet for image recognition purposes. Nevertheless, it should be noted, that REL-PCANet contains just ***some*** parts of discussed instruments and it is specified for IDI estimation.

### *3.2 Relative Attributes Rating Model theoretical framework*

RELARM contains 3 stages:

1. Normalization of input data – unification of initial model parameters for their comparison using linear scaling method.
2. Obtaining of the relative attribute ranking functions values:
    a. calculation of relative PCA attribute ranking functions;
    b. mapping of normalized parameter vector in the space of relative PCA attribute ranking function values;
    c. formation of the rating vector.
3. Application of k-means clustering algorithm for obtainment of results.

**Normalization.** Suppose that our model consists of $N$ factors and $M$ objects. We apply a linear scaling method (min-max transformation) in order to standardize rating model parameters for their comparability.

Let $p_{ij}$, $i \in [M], j \in [N]$ denote the initial value of the *j-th* parameter of the *i-th* rating object. We define a normalized value $b_{ij}$ of $p_{ij}$, where $i \in [M], j \in [N]$, depending on the *j-th* factor's influence on the model property studied.

If an increase of $p_{ij}$ index value has a positive impact on the final analyzed property, the formula becomes:

$$b_{ij} = \frac{p_{ij} - \min_i p_{ij}}{\max_i p_{ij} - \min_i p_{ij}}, i \in [M], j \in [N]. \tag{1}$$

If a model parameter increase has a negative effect on the final rating, then normalized value $b_{ij}$ is calculated as:

$$b_{ij} = \frac{\max_i p_{ij} - p_{ij}}{\max_i p_{ij} - \min_i p_{ij}}, i \in [M], j \in [N]. \tag{2}$$

As a result, each object is described by a $(1 \times N)$ dimension row vector of normalized parameters:

$$b_i^T = (b_{i1}, \dots, b_{iN}) \in [0,1]^N, i \in [M]. \tag{3}$$

Let

$$B := \{b_i\}, i \in [M]. \tag{4}$$

denote a set of normalized parameters.

**Obtainment of relative attribute ranking functions values.** The *p-th* relative PCA attribute of vector $b_i \in B$ $i = 1,2,...M$ is a vector $A_{ip}$:

$$A_{ip} = (b_{i1}w_{1p}, b_{i2}w_{2p},..., b_{iN}w_{Np}), \quad p = 1....,N. \tag{5}$$

Where

$$w_k = \begin{pmatrix} w_{1k} \\ \vdots \\ w_{Nk} \end{pmatrix} \tag{6}$$

denotes $l_1$-normalized PCA components of the set B (3.4) with principal component variances $\lambda_k, \lambda_1 \geq \lambda_2 \geq \cdots \geq \lambda_N$.

Here in accordance with the concept of relative attributes the *p-th* main attribute has a stronger presence in vector $b_i$ than in vector $b_j$, if $l_1$-norm of vector $A_{ip}$ is greater than $l_1$-norm of vector $A_{jp}$:

$$\sum_{k=1}^{N} b_{ik}|w_{kp}| \geq \sum_{k=1}^{N} b_{jk}|w_{kp}| \tag{7}$$

The *ranking vector* for *p-th* main attribute is the vector $\widetilde{w}_p^T$:

$$\widetilde{w}_p^T = (|w_{1p}|,...|w_{Np}|), \tag{8}$$

and the *ranking function* is defined by formula:

$$r_p(b_i) = b_i^T \widetilde{w}_p \tag{9}$$

([16]).

Also the $N \times d$ matrix $W$ is defined as:

$$W = \begin{pmatrix} |w_{11}| & \cdots & |w_{1d}| \\ \vdots & \ddots & \vdots \\ |w_{N1}| & \cdots & |w_{Nd}| \end{pmatrix}, \tag{10}$$

where the number of principal components $d$ is determined to avoid the influence of «data noise».

The **rating vector** $\Lambda$ is defined as:

$$\Lambda := (\lambda_1, \lambda_2, ..., \lambda_d). \tag{11}$$

Let $f: B \to \mathrm{R}^d$ be a map of the set B to the space $\mathrm{R}^d$ of relative attribute ranking functions values defined by formula:

$$a_i^T := f(b_i^T) = b_i^T \times W. \tag{12}$$

Here:

$$a_i^T = (r_1(b_i),\ r_2(b_i),\ ...,\ r_d(b_i)) = (\sum_{k=1}^{N} b_{ik}|w_{k1}|, \sum_{k=1}^{N} b_{ik}|w_{k2}|, ..., \sum_{k=1}^{N} b_{ik}|w_{kd}|) \tag{13}$$

For the *i-th* rating object each component of vector $a_i^T$ indicates the degree of influence of object's parameter changes with respect to the corresponding principal component.

**Application of k-means clustering algorithm for obtainment of results.** After relative attribute ranking functions values are obtained it becomes the time for partitioning objects into classes. Classification implies the following actions:

1. Application of k-means clustering algorithm application to obtained vectors of relative PCA attribute ranking function values (13) .
2. Projection of Cluster centers on the rating vector $\Lambda$ (11). The output cluster centers we denote by $CC_q, q \in (1,2,...,k)$. Module of projection of the *q-th* cluster center on the rating vector $\Lambda$ (11) is calculated as follows:

$$PR_q = |(CC_q, \Lambda)|, q \in (1,2,...,k). \tag{14}$$

3. Ranking of centers projection on the rating vector in descending order.

Here we determine the importance of clusters by projection of their centers on the rating vector.

*3.3 RankNet theoretical framework*

RankNet is a pairwise approach and it applies neural network for its construction. It is a feedforward network with a single output neuron. RankNet uses objects features as initial data and with the stochastic gradient decent back-propagation algorithm it trains the weights, bias to perform the output value. Basically, RankNet is trained on pairs of initial vectors where each of them has special label. And as the result it gives a real number out of the initial *feature* vector [2].

Here we briefly describe RankNet algorithm in the context of our work.

Suppose we have a set of feature vectors

$$D = \{a_i\}, \tag{15}$$

where $a_i \in R^d, i = (1, \ldots M)$. The inputs of the network are pairs of vectors of the relative attribute ranking functions:

$$\{a_i, a_j\}, \ a_i, a_j \in D. \tag{16}$$

Usually the desired outputs or targets have to be presented for network training and in the case of RankNet they have to be presented by a probabilities matrix $T_{Rnet}$:

$$T_{Rnet} = (t_{ij}), i,j = (1, \ldots M), \tag{17}$$

Where element $t_{ij}$ show the probability of feature vector $a_i$ having higher grade than $a_j$.

RankNet is used to find a special ranking function $f: \mathbb{R}^d \mapsto \mathbb{R}$ which gives scores to feature vectors. Here $f(a_i) > f(a_j))$ means that $a_i$ has higher ranking than $a_j$ with posterior probability $P_{ij}$ which is calculated using the following formula:

$$P_{ij} \coloneqq \frac{1}{1+e^{-(rank_i - rank_j)}}, \tag{18}$$

where $rank_i = f(a_i), rank_j = f(a_j)$.

RankNet employs cross entropy function for network performance evaluation in the following way:

$$CE_{ij} \coloneqq -t_{ij} \log(P_{ij}) - (1 - t_{ij}) \log(1 - P_{ij}). \tag{19}$$

RankNet adjusts weights in the network using gradient descent backpropagation to minimize the loss value.

*3.4 Ranking Relative Principal Component Attributes Network Model (REL-PCANet)*

Here we present our proposed approach for estimation of the Inclusive Development Index – the Ranking Relative Principal Component Attributes Network Model (REL-PCANet). For that purpose, we use RELARM and Deep relative attributes constructions, however adjusting them in correspondence with our needs. The REL-PCANet takes into account dynamics of countries' changes and reflects it in final ranking.

The Ranking Relative Principal Component Attributes Network Model contains 3 stages:

1. Application of RELARM to initial IDI data;

2. New approach for formation of probabilities matrix $T_{Rnet}$;

3. Application of REL-PCANet with obtainment of final rankings.

*3.4.1 Application of RELARM*

The main goal of this phase is to obtain relative attribute ranking functions values $a_i^T$ (13) and the matrix $T_{Rnet}$ (17) with elements $t_{ij}$. Firstly, we take 12 variables for IDI calculation used by the World Economic Forum (Table 1) and normalize them based on the parameter influence on the final result using formulas (1) and (2):

Table 1. Variables for IDI calculation

| Positive influence | GDP per capita, $ | Wealth gini | Negative influence | Poverty rate, % |
|---|---|---|---|---|
| | Labor productivity, $ | Median income, $ | | Carbon intensity, kg per$ of GDP |
| | Healthy life expectancy, yrs | Adjusted net savings, % | | Public debt, % |
| | Employment, % | Net income, gini | | Dependency ratio, % |

It is important to note that RELARM should be calculated separately for advanced and emerging countries as the final IDI ranking is presented for these two groups.

After normalization, if some data is missing it is necessary to complement it to avoid outliers. Here it can be realized on the principle of income thresholds presented by WEF in [18]. Assuming there are 4 groups of countries (advanced economies, upper-middle income economies, lower-middle income economies and low income economies) if occurs missing data for some of country's variable then we take the average group's value for this particular parameter. **Note.** We are not filling empty cells with zeros because it might cause outliers in the model's output (and also because the real value is not zero, it is just lack of resources to find it).

After the normalized set of parameters $B$ (4) is obtained, we find the rating vector (11) and relative PCA attribute ranking function values (13) using procedure described in 3.1. For matrix $W$ (10) we recommend to take the number of principle components $d$ providing approximately 95% of data information.

Next, we run the third stage of RELARM – application of k-means clustering algorithm. Here we apply the classifying algorithm for obtaining clusters needed for formation of matrix $T_{Rnet}$. Considering that there are 30 advanced countries in IDI and 77 emerging we suppose to group them in 5 clusters. The reason is that the higher of clusters number can raise not desirable movements (as k-means clustering algorithm has certain downsides) which have no reasonable economic basis. Also, in order to ensure stable algorithm results, one should perform several iterations before forming final clusters, so that we suggest running 50 and more iterations. Here we applied the classifying algorithm to vectors (13) for obtaining clusters needed for the next step - formation of matrix $T_{Rnet}$.

### 3.4.2 New approach for formation of the target probabilities matrix $T_{Rnet}$

The special emphasis in the Ranking Relative Principal Component Attributes Network is given to the probabilities matrix $T_{Rnet}$. We construct it in the way to reflect countries' movements between clusters from year to year and therefore adjusting final IDI rankings according to these quality changes. Additionally, in our framework matrix $T_{Rnet}$ is intended to catch upward and downward countries' trends. Furthermore, matrix $T_{Rnet}$ is based on the clusters obtained after realization of RELARM. Suppose that countries are divided into 5 clusters.

Here, we divide calculation principles of matrix $T_{Rnet}$ into two categories: for the first year of IDI and the following years.

**A. Matrix $T_{Rnet}$ for the 1st year of IDI calculation.** Here we assume that element $t_{ij}$ of matrix $T_{Rnet}$ can take one of the following 8 values $t_{ij} \in \{0, 0.35, 0.4, 0.45, 0.55, 0.6, 0.65, 1\}$. There can be distinguished 3 cases for defining the value of $t_{ij}$:

*A1.* Country $i$ lies in cluster with <u>higher</u> value of cluster center projection on the rating vector than country $j$: $t_{ij} = 1$.

*A2.* Country $i$ lies in cluster with <u>lower</u> value of cluster center projection on the rating vector than country $j$: $t_{ij} = 0$.

*A3.* Country $i$ lies in the <u>same</u> cluster as country $j$ then we calculate projections of countries' relative attribute ranking functions values on the rating vector Λ (3.11) and also find rankings of countries within every cluster. Here the higher projection value the higher ranking. Suppose that cluster $L$ has $Q$ countries and ranking of each country is denoted as $r_p$ where $p \in \{1, ..., Q\}$. So, if projection of the $i^{th}$ country's relative attribute ranking functions value $a_i^T$ on the rating vector $\Lambda$ is higher than projection of $a_j^T$ on $\Lambda$ then for $r_i = r_j - 1$, $t_{ij} = 0.55$, for $r_i = r_j - 2$, $t_{ij} = 0.6$, for $r_i \geq r_j - 3$, $t_{ij} = 0.65$. And if projection of the $i^{th}$ country's relative attribute ranking functions value $a_i^T$ on the rating vector $\Lambda$ is lower than projection of $a_j^T$ on $\Lambda$ then for $r_i = r_j + 1$, $t_{ij} = 0.45$, for $r_i = r_j + 2$, $t_{ij} = 0.4$, for $r_i \geq r_j + 3$, $t_{ij} = 0.35$.

**B. Matrix $T_{Rnet}$ for the 2$^{nd}$ year of IDI calculation.** Starting from the second year of the IDI estimation the goal is to take into account countries' movements between clusters if such occur. That is why $t_{ij}$ now can take one of the 9 values $t_{ij} \in \{0, 0.35, 0.4, 0.45, 0.5, 0.55, 0.6, 0.65, 1\}$. Here if country $i$ does not change its cluster category in the studied year, $t_{ij}$ is calculated according to the algorithm described in paragraph A. However, if country upgrades or downgrades from the previous year cluster there can be distinguished the following cases. For better understanding assume that there are clusters A (the highest), B and C (the lowest).

**B1.** Country $j$ downgraded from its cluster (let's say from A to B). Then for each country $i$ (that stayed in B) which has higher projection value of $a_i^T$ on the rating vector $\Lambda$ than $a_j^T$ does, $t_{ij} = 0.65$. If it has lower projection value then situation is uncertain and $t_{ij} = 0.5$. Situation for when country $j$ upgraded from its cluster is vice versa.

**B2.** Countries $i$ and $j$ both downgraded or upgraded to a particular cluster. Then $t_{ij} = 0.5$ because still their interconnections are indefinite.

**B3.** Country $j$ upgraded from C to B and country $i$ downgraded from A to B. If $i$ has higher projection value of $a_i^T$ on the rating vector $\Lambda$ then $t_{ij} = 0.5$ and if lower - $t_{ij} = 0.65$. Situation for when country $j$ moved down and $i$ moved up from their clusters is vice versa.

**Note.** We only specified estimation of matrix $T_{Rnet}$ for 2 years to show how it can possibly take into account changes from the previous years. However, in practice the rule for the following periods can be established in accordance with any economic reasons of assessing organization.

As the result of 2 stages of the Ranking Relative Principal Component Attributes Network we obtained feature vectors $a_i^T, i \in (1, ..., M)$ and matrix $T_{Rnet}$ of target probabilities.

### 3.4.3 The REL-PCANet framework

***The Ranking Relative Principal Component Attributes Network Model*** architecture is presented on Figure 1. It contains 4 hidden layers with a single output neuron and is divided into two sections: deep feature extraction and ranking parts. We use log-sigmoid function as transfer function for the network and crossentropy function (19) as a loss/performance function as described in 3.3.

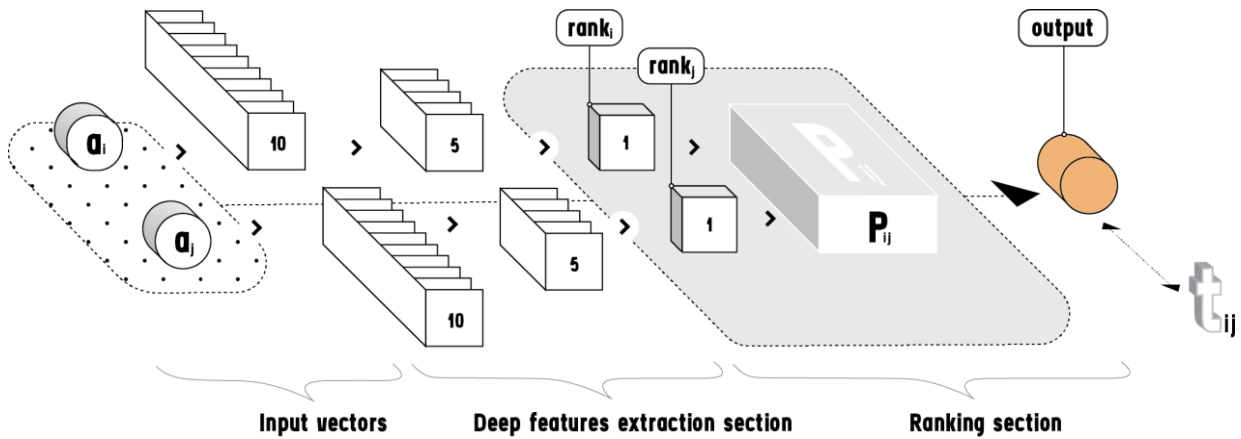

Figure 1. The REL-PCANet architecture (numbers on the figure denote the number of nodes)

**Deep features extraction section.** It starts from input vectors which are relative attribute ranking functions values (16) presented to the network in pairs like it was described in 3.3 with their target probabilities $t_{ij}$ (17). It contains 3 hidden parallel layers. We call this section «deep feature extraction» as input vectors already went through RELARM process helped to distinguish relevant characteristics of initial IDI data in the form of relative attributes and this part of network further deepens the procedure of finding features.

**Ranking section.** That last layer of the REL-PCANet is based on RankNet that is why it is called the ranking section. We find $rank_i$ and $rank_j$ values within the ranking layer via formula:

$$rank_i \coloneqq weight^T x_i + bias, \qquad (20)$$

Where $x_i$ is a feature vector obtained in network, $weight$ and $bias$ are the layer's weights and bias.

As we have values of $rank_i$ and $rank_j$ it is possible to calculate $P_{ij}$ (18) and therefore the network's loss $CE_{ij}$ by formula (19). After the network is trained using resilent backpropagation alghorithm until its loss is minimized.

Finally, after the network is trained we receive scores for each country and normalize them using linear transformation in the scale from 1 to 7 – the World Economic Forum currently uses that range of values for IDI estimation. The rankings can be obtained by putting the normalized values in descending order – the higher the value the better ranking. We used a simple way to normalize the obtained scores just to be able to compare the results on year to year basis, however, it might be a subject for further discussions.

## IV. Empirical study

In this part we show results obtained by the Ranking Relative Principal Component Attributes Network Model and compare them with the Inclusive Development Index scores and rankings. For calculation we used data of 12 variables presented in the World Economic Forum reports for the years 2017 [18] and 2018 [19]. In addition, in this study we compare results for the group of 29 advanced countries (however the REL-PCANet is computed for 30 countries, including Singapore for which WEF presented data but did not calculate IDI). We performed REL-PCANet based on algorithm described in Section III. Obtained results for the IDI estimated by the REL-PCANet are shown in Figure 2 (green narrows show that country obtained higher rating than in previous year and the reds vice versa).

| Economy | 2018 Score | 2018 Rank | 2017 Rank |
|---|---|---|---|
| Luxembourg | 7.00 | 1 | 1 |
| Singapore | 6.75 | ↑ 2 | 4 |
| Norway | 6.72 | ↓ 3 | 2 |
| Switzerland | 6.53 | ↓ 4 | 3 |
| Denmark | 5.98 | 5 | 5 |
| Iceland | 5.68 | 6 | 6 |
| Austria | 5.45 | ↑ 7 | 8 |
| Sweden | 5.00 | ↓ 8 | 7 |
| Netherlands | 4.89 | 9 | 9 |
| New Zealand | 4.53 | ↑ 10 | 11 |
| Canada | 4.53 | ↑ 11 | 13 |
| Ireland | 4.52 | ↓ 12 | 10 |
| Korea, Rep. | 3.82 | ↑ 13 | 16 |
| United States | 3.74 | 14 | 14 |
| Australia | 3.70 | 15 | 15 |
| Germany | 3.41 | ↓ 16 | 12 |
| United Kingdom | 3.08 | ↑ 17 | 18 |
| France | 2.68 | ↑ 18 | 19 |
| Finland | 2.68 | ↓ 19 | 17 |
| Belgium | 1.84 | 20 | 20 |
| Czech Republic | 1.80 | 21 | 21 |
| Spain | 1.70 | 22 | 22 |
| Israel | 1.54 | 23 | 23 |
| Italy | 1.53 | ↑ 24 | 25 |
| Estonia | 1.39 | ↑ 25 | 26 |
| Slovenia | 1.38 | ↓ 26 | 24 |
| Portugal | 1.35 | 27 | 27 |
| Japan | 1.22 | ↑ 28 | 29 |
| Slovak Republic | 1.08 | ↓ 29 | 28 |
| Greece | 1.00 | 30 | 30 |

Figure 2. Ranking Relative Principal Component Attributes Network results for 30 advanced economies

The whole changes in both scores and rankings are presented in Figure 3.

We can see that Ranking Relative Principal Component Attributes Network gives robust adequate results both for scores and rankings, especially due to specific formation of probabilities matrix $T_{Rnet}$. Additionally, REL-PCANet takes into account dynamic changes of countries' inclusive development which in whole does not exceed 2-3 steps from the rank of 2017 and also does not perform any dramatic changes in scores (average value of change equals to -0.2). However, it should be noted that as the

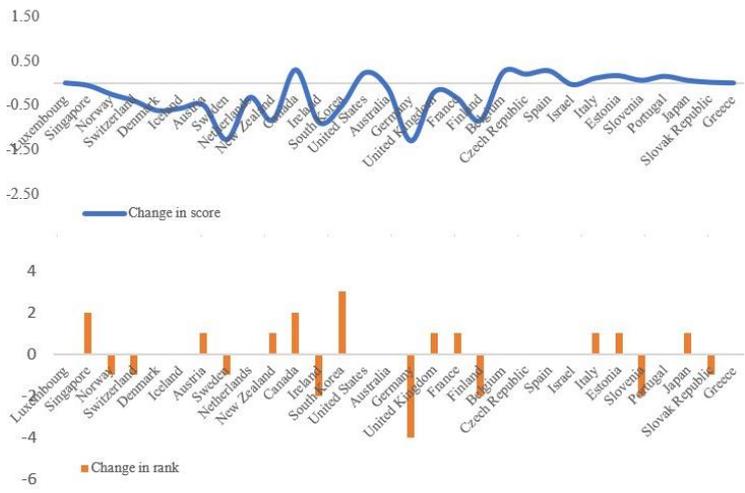

Figure 3. Changes in scores and ranks REL-PCANet 2018

Ranking Relative Principal Component Attributes Network takes into account countries' interdependencies and each period performs ranking according to the system for that specific time there might occur movements of country's score value and ranking with opposite directions. In such cases, one should analyze precisely reasons for such country changes: if the inclusive development enhanced or deteriorated because of internal country's problems or as a result of overall upward/downward trends in advanced economies.

Next, we compare our results of IDI scores and rankings for 2 years presented by WEF. Figure 4 reflects the differences. Here we can conclude that rankings of REL-PCANet and WEF do not look alike and in average the distance between them for 2017 values amounts to 3 points and for 2018 it is raised to almost 4 points (which also shows robustness of REL-PCANet). However, we also compared scores obtained for 2 models. In order to perform that, we took REL-PCANet scores for 2018 and normalized them according to minimum and maximum value of the IDI 2018. Results are presented on Figure 5. As is can be seen overall scores have similar values, the average difference in scores does not exceed 0.5 points and only 4 countries have 1 point higher score in WEF ranking then in REL-PCANet. Additionally, we performed another test to analyze how the probabilities matrix $T_{Rnet}$ reflects dynamic changes. So that, we calculated matrix $T_{Rnet}$ for 2018 based on principles for 2017 (not taking into account countries' movements between clusters), run the REL-PCANet algorithm and normalized obtained scores according to minimum and maximum values of WEF IDI 2018. After, we compared these scores with the WEF results and REL-PCANet estimations based on matrix $T_{Rnet}$ calculated in conjunction with cluster changes. Figure 6 reflects analysis outcome. It can be noticed that REL-PCANet scores based on dynamic matrix $T_{Rnet}$ for most countries have larger difference with the WEF results than REL-PCANet scores based on the not dynamic $T_{Rnet}$. That might reveal that IDI estimated by WEF does not reflect previous countries' conditions while REL-PCANet takes into account deep interdependencies as well as occurred changes.

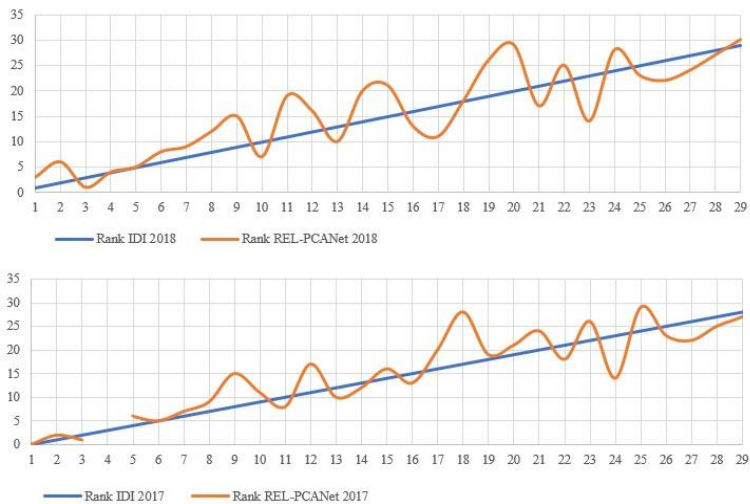

Figure 4. Comparison of ranking results IDI and REL-PCANet 2017-2018

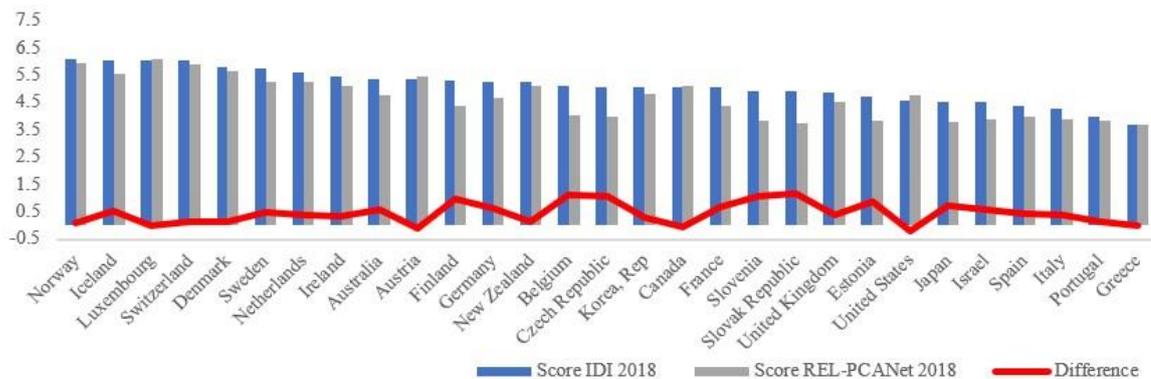

Figure 5. Comparison of score results IDI and REL-PCANet 2017-2018

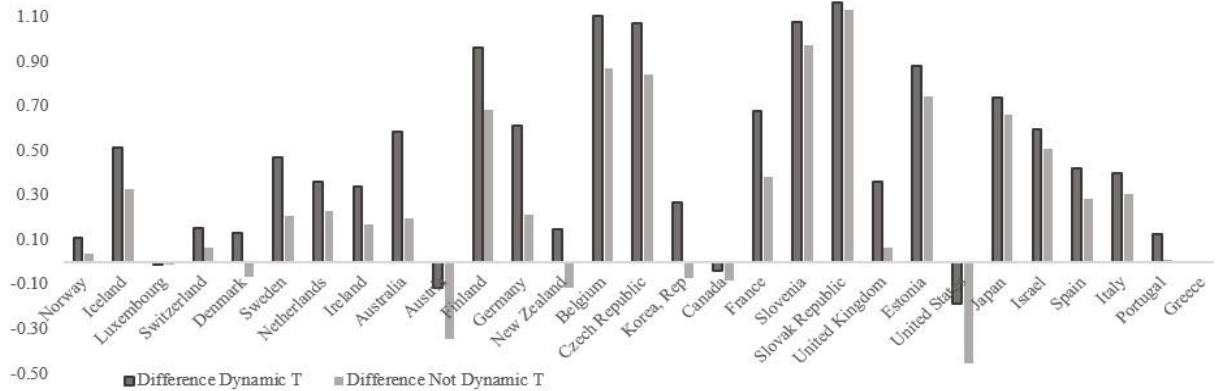

Figure 6. Difference between WEF score and scores of REL-PCANet estimated by dynamic and not dynamic matrix T, 2018

Overall analysis showed that REL-PCANet:

— ensures robust and adequate results and probabilities matrix $T_{Rnet}$ can work as a tool for reflection of dynamic changes of countries' inclusive development;
— assigns similar to WEF scores with a little difference;
— performs slightly different ranking results in comparison with WEF.

Therefore, the Ranking Relative Principal Component Attributes Network Model proved that it can be applied for the Inclusive Development Index estimation. It is based on reliable and transparent methods for calculation that is why obtained rankings can be assumed as more accurate.

### V. Conclusion

We proposed a novel approach for estimation of the World Economic Forum's Inclusive Development Index - the Ranking Relative Principal Component Attributes Network Model (REL-PCANet).

Study showed that REL-PCANet reflects of countries' year to year changes in final scores and rankings due to special construction of target probabilities matrix $T_{Rnet}$ suggested for REL-PCANet. Model tests proved that model reflects deep countries' interdependencies as well as ensures robust results.

Moreover, comparison of the REL-PCANet and WEF IDI scores and rankings revealed that REL-PCANet are just slightly different from the WEF's. Also, analysis showed great relevance of the REL-PCANet results and stressed the problem of taking into account the dynamics of country rankings over time in existing estimations.

In conclusion, the Ranking Relative Principal Component Attributes Network Model proved to be a reliable, transparent and accurate ranking system for IDI estimation. It can be recommended for practical implementation in the WEF's framework. Also, it's results can be taken as a base for countries' reforms for enhancement of their inclusive development.